# Chirality in non-Hermitian photonics

**Sunkyu Yu, Xianji Piao, and Namkyoo Park\***

*Photonic Systems Laboratory, Dept. of Electrical and Computer Engineering,*
*Seoul National University, Seoul 08826, Korea*
*\*Corresponding author:nkpark@snu.ac.kr*

Chirality is ubiquitous from microscopic to macroscopic phenomena in physics and biology, such as fermionic interactions and DNA duplication. In photonics, chirality has traditionally represented differentiated optical responses for right and left circular polarizations. This definition of optical chirality in the polarization domain includes handedness-dependent phase velocities or optical absorption inside chiral media, which enable polarimetry for measuring the material concentration and circular dichroism spectroscopy for sensing biological or chemical enantiomers. Recently, the emerging field of non-Hermitian photonics, which explores exotic phenomena in gain or loss media, has provided a new viewpoint on chirality in photonics that is not restricted to the traditional polarization domain but is extended to other physical quantities such as the orbital angular momentum, propagation direction, and system parameter space. Here, we introduce recent milestones in chiral light-matter interactions in non-Hermitian photonics and show an enhanced degree of design freedom in photonic devices for spin and orbital angular momenta, directionality, and asymmetric modal conversion.

*Keywords:* Chirality, Non-Hermitian Photonics, Parity-Time Symmetry, Exceptional Point

*OCIS codes:* (080.6755) Systems with special symmetry; (160.1585) Chiral media; (000.1600) Classical and quantum physics; (260.6042) Singular optics; (260.5430) Polarization

# I. INTRODUCTION

Chirality, the asymmetry between mirror images also known as enantiomers, is universally observed in nature. The study of chiral phenomena has been a critical issue in physics, biology, and chemistry because of the theoretical importance as shown in weak interactions between fermions [1] and the practical applications using the handedness of physical objects such as chemical reactions between enantiomers [2]. Chirality has also played a critical role in photonics after the discovery of optical rotation by Arago in 1811 [3]. The well-accepted definition of chirality in photonics is the differentiated responses of "optical polarization enantiomers" [4,5], which depict right- (R-) and left- (L-) circular polarizations (CPs) that correspond to the spin angular momentum (SAM) of photons. Traditional chiral responses of light include (i) CP-differentiated phase velocities inside three-dimensional (3D) chiral materials [6], and (ii) CP-differentiated optical absorption through 2D chiral materials, also known as circular dichroism [7]. This optical chirality has been the foundation of various functionalities both in classical optics and nanophotonics, such as optical sensing of enantiomers [8], negative refractions [9], control and utilization of photonic angular momenta [10], photonic topological insulators [11], and plasmonic color generations [12].

Recent progress in the emerging field of non-Hermitian photonics [13-22] inspired by the concept of parity-time (PT) symmetry [23,24] has provided a new perspective on optical chirality. Since the discovery of real spectra in PT-symmetric potentials despite their non-Hermitian Hamiltonians [23,24], non-Hermitian photonics that explores photonic systems composed of gain and loss media has attracted much attention, focusing on the classical simulation of PT-symmetric quantum systems [25,26] and the application of PT-symmetric phenomena to photonic devices [27-29]. Most intriguing phenomena in non-Hermitian photonic systems are observed near the exceptional point (EP) [15], which indicates the transition between real and complex eigenvalues and the coalescence of eigenmodes. In particular, as first described by Heiss *et al.* [30-33], the chirality observed in the EP is not the platform-specific phenomenon, such as the mixing of electric and magnetic responses in traditional optical chirality [6], but it is the general symmetry-protected phenomenon in a non-Hermitian system. This platform-invariant nature of EP-induced chirality inspired the extension of the definition of optical chirality to physical quantities other than polarizations: orbital angular momentum (OAM) for structured light, canonical momentum for wave propagation, and the trajectory of optical states in a system parameter space. Although several recent articles have reviewed the significant milestones in non-Hermitian photonics [13-17], a comprehensive review of chiral phenomena in non-Hermitian photonics is still lacking.

In this review, we introduce recent achievements in non-Hermitian photonics and related fields that have provided a new path to wave chirality, in particular, focusing on practical device applications. We discuss the observation of chirality in different physical domains (Fig. 1): the traditional polarization domain for the SAM of light (Section II), beam wavefront for the OAM of light (Section III), propagation direction defined by canonical momentum (Section IV), and state evolution in the geometry of a material parameter space (Section V), which can also be classified by chirality for real-space quantities (Sections II-IV) and abstract system parameters (Section V). We show that a novel degree of freedom obtained by extending the scope of optical chirality to complex potentials has enabled the realization of a "singular handedness material" for various quantities of light, achieving universal light polarizers, structured light lasing, and directional photonic devices.

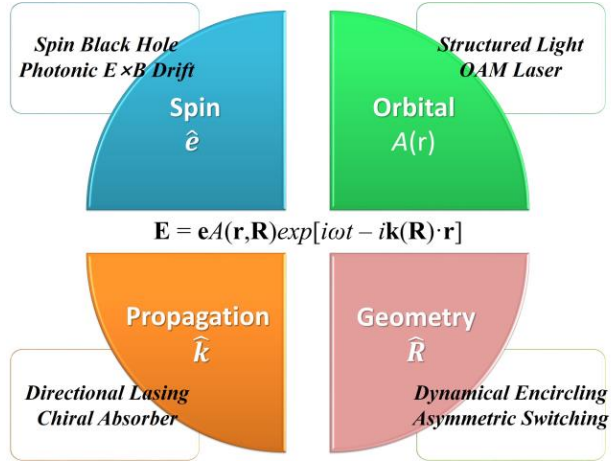

Figure 1. Chirality in different physical domains of non-Hermitian photonics. For the optical field, $\mathbf{E} = \mathbf{e}A(\mathbf{r},\mathbf{R})\exp(i\omega t - i\mathbf{k}(\mathbf{R})\cdot\mathbf{r})$, where $\mathbf{R}$ is the system parameter vector and $\mathbf{r}$ is the position vector, the extended definition of optical chirality in non-Hermitian photonics can be classified according to each physical quantity: polarization $\mathbf{e}$ for SAM, wavefront $A(\mathbf{r},\mathbf{R})$ for OAM, canonical momentum $\mathbf{k}(\mathbf{R})$ for wave propagation, and the geometry of state evolution in the system parameter space $\mathbf{R}$. The system parameter $\mathbf{R}$ represents the complex optical potential that determines the condition of PT symmetry, including on-site and hopping constants defined by structural and material parameters.

## II. CHIRAL POLARIZATIONS

Since the discovery of PT symmetry by Bender [23], this symmetry has been well accepted as an important theoretical tool for achieving real eigenvalues in complex potentials with non-Hermitian Hamiltonians. PT-symmetric potentials also support regimes in which some eigenvalues are complex, but the most intriguing phenomena occur near the transition point between real and complex phases, which is called the EP [15]. One of the fascinating features of the EP is the chirality of eigenmodes observed by Heiss [30,31]. To illustrate this novel type of chirality in non-Hermitian systems, we start from a simple two-level PT-symmetric system, which describes weakly coupled elements that directly correspond to the polarization space with two degrees of freedom, for example, $x$- and $y$-linear polarizations. The system is described by the following 2D equation [25,26,34]:

$$i\frac{d}{d\xi}\begin{bmatrix}\psi_G\\\psi_L\end{bmatrix}=\begin{bmatrix}V_r+iV_i & \kappa\\ \kappa & V_r-iV_i\end{bmatrix}\begin{bmatrix}\psi_G\\\psi_L\end{bmatrix}, \quad (1)$$

where $V_r$ and $V_i$ are the real and imaginary parts of the on-site potentials, respectively, $\kappa$ is the real-valued coupling between two states with the convention of $\kappa > 0$, $\psi_{G,L}$ denotes the field of each element, and $\xi$ represents the physical axis for the state evolution, which can be a time ($t$) or spatial axis ($z$). The steady-state solution with $d\Psi/d\xi = -i\gamma\Psi$ for $\Psi = (\psi_G,\psi_L)^T$ then leads to complex eigenvalues $\gamma_{1,2}$ and the corresponding eigenmode $v_{1,2}$, as

$$\gamma_{1,2} = V_r \pm \sqrt{\kappa^2 - V_i^2}, \quad (2)$$

$$v_{1,2} \sim \begin{bmatrix}\kappa\\ -iV_i \pm \sqrt{\kappa^2 - V_i^2}\end{bmatrix}. \quad (3)$$

At the EP ($\kappa = V_i$) that determines the transition between real and complex eigenvalues across $\gamma_1 = \gamma_2 = V_r$, Eqs. (2) and (3) show that the coalesced eigenmode $v_1 = v_2$ has the chiral form of $(1, -i)^T$. The given 2D non-Hermitian system is thus chiral at the EP because the system possesses only a single eigenmode with the chiral form $(1, -i)^T$, showing broken mirror symmetry. We also note that the non-Hermitian system at the EP has a geometric multiplicity of 1, which represents a reduction in the system dimensionality.

Because of the platform-transparent form of Eq. (1), the described chirality protected by PT symmetry is universally accessible in any non-Hermitian system with two basis vectors, such as the polarization domain. The emergence of PT-symmetry-protected chirality is described in Fig. 2a-e [35,36], showing the evolution of the eigenpolarization states according to the phase of PT symmetry. As shown, the PT-symmetric system can possess a singular handedness polarization at the EP that corresponds to the pure SAM (Fig. 2c) when compared to two elliptic nonorthogonal polarizations in non-Hermitian potentials around the EP (Fig. 2b,d,e) and two orthogonal polarizations in Hermitian potentials (Fig. 2a). Due to the low dimensionality with the singular existence of a chiral eigenmode, the given PT-symmetric system also represents the "customization" of optical states into the chiral form. This molding of optical states results in the convergence of optical polarization to the specific SAM, similar to the spin black hole behavior [36], which enables the universal circular polarizers, that is, the achievement of chiral light irrespective of the incident polarization state (Fig. 2f). The PT-symmetry-protected chiral polarization and polarization convergence have been demonstrated experimentally in lattice structures (Fig. 2g) [36], photonic molecules (Fig. 2h) [35] in the THz regime, and numerically in metasurfaces [37,38].

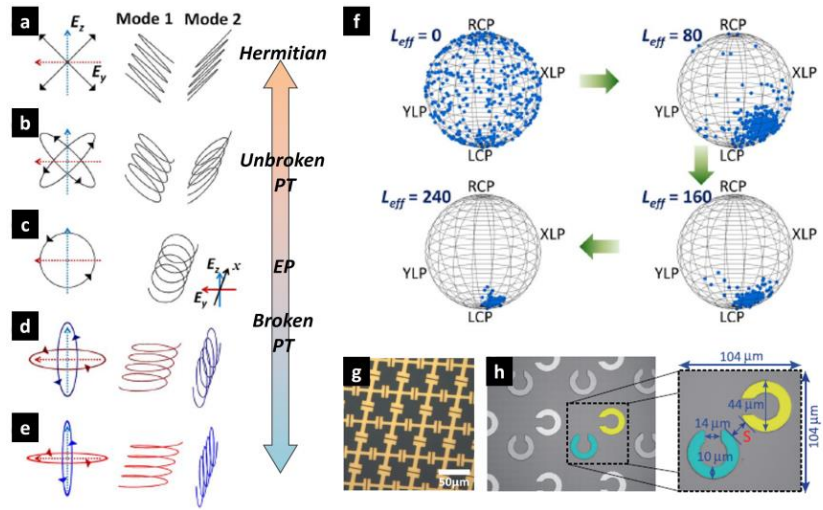

Figure 2. Chiral polarizations in non-Hermitian potentials. (a-e) Evolution of eigenpolarizations according to the phase of PT symmetry [36]: (a) Hermitian, (b) unbroken, (c) EP, and (d,e) non-Hermitian states. (f) The convergence of polarizations to the LCP state, showing spin black hole [36]. (g,h) The experimental platform for PT-symmetry-protected chirality: (g) lattice structures [36] and (h) photonic molecules [35]. Figure adapted from (a-g), ref. [36], OSA; and (h), ref. [35], APS, with permission.

The theory of chiral polarization at the EP has been generalized to the analogy of relativistic electrodynamics [39,40], the arbitrary control of polarization states [41,42], and the realization of a transversely zero SAM [43]. Starting from the similarity between Maxwell's equations for polarization states and the Lorentz force equation for relativistic particle movements [44], the understanding of chiral polarization at the EP as the E×B polarization drift has been demonstrated [39]. This generalization has enabled the realizations of the directional operation of linear polarizers [39] and the stable or sensitive qubit evolution in non-Hermitian systems depending on the phase of PT symmetry [40]. The degrees of freedom in non-Hermitian systems have also been applied to the arbitrary control of polarization states for their sensitive detection [41], and the control of polarization states using the near-thresholdless EP in plasmonic resonators [42]. Furthermore, as shown in the abstract form of Eq. (1), the PT-symmetric transformation of the linear basis to a chiral polarization in Fig. 2 can be easily extended to the transformation of the chiral basis to a linear polarization, which has been applied to the annihilation of the transverse spin, first demonstrating the linear polarization in the meridional plane of guided waves [43]. As shown, the introduction of chirality with non-Hermitian photonics has provided a new degree of freedom for the precise control of the 3D polarization states of light, achieving practical devices of universal circular polarizers, directional linear polarizers, polarization converters, and stable or sensitive qubit control in non-Hermitian quantum systems.

## III. CHIRAL WAVEFRONT

As shown in the abstract form of Eq. (1), the chirality in PT-symmetric systems is platform-transparent and is protected by PT symmetry in contrast to the optically specific form of traditional optical chirality, that is, the mixing of electric and magnetic responses [6]. Therefore, by assigning a proper optical quantity to the basis of PT-symmetric systems, a new type of optical chirality for the target quantity has been achieved, such as the chirality of wavefronts and directionality. In this section, we summarize the recent achievements in the "chiral wavefront" of light obtained from non-Hermitian optical potentials, which provides the OAM of light, also known as structured light [45].

The OAM is a fundamental property of photons, originating from the singularity in the wavefront of light defined by the topological charge. Because the OAM order is unbounded in contrast to two degrees of freedom in the SAM, access to the OAM of light has been desired to increase the information capacity in optical communications [46]. To generate the necessary phase evolution in the beam wavefront, recent methods using Hermitian potentials adopt the conversion of SAM to OAM [47] using pattered layers [48], spatiotemporally modulated microcavities [49], or optically controlled interactions in photonic molecules [50]. The chiral nature of non-Hermitian systems at the EP has also provided additional degrees of freedom in generating the OAM by constructing the unidirectional phase evolution in the wavefront.

Compared to the breaking of the mirror symmetry between circular polarizations, chiral wavefronts have been achieved from the breaking of the mirror symmetry between circular propagations of light, namely, clockwise (CW) and counterclockwise (CCW) wave propagation, initially studied in dissipative resonators [51]. This passive realization of the EP has been extended to the realization of exact PT symmetry using active materials, which was applied to construct a vortex laser beam with a nontrivial OAM [52] (Fig. 3a-c). In this system, the condition of PT symmetry is achieved with alternating Cr/Ge bilayer and Ge single-layer structures periodically implemented on the active layer, which is composed of an InP substrate with a top layer made up of an InGaAsP multiple quantum well (Fig. 3a). The number of alternating structures selectively breaks the mirror symmetry of the target azimuthal number of the whispering gallery mode, and at the EP, the CCW mode unidirectionally circulates the resonator. This chiral eigenmode generates the laser beam with the chiral wavefront (Fig. 3b) by introducing the sidewall scatterers to couple the lasing emission upward (Fig. 3c). In a recent work [53], the generation of the OAM using non-Hermitian potentials has been realized with the S-shaped optical structure with tapering, which allows a broadband response and the stable generation of the OAM.

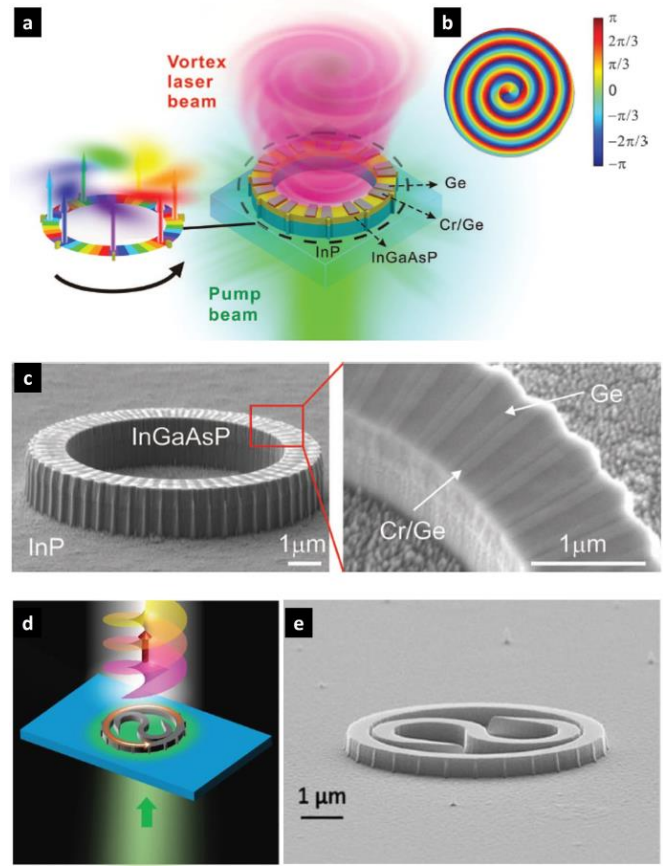

Figure 3. Chiral wavefronts in non-Hermitian potentials. (a-c) OAM microlaser [52]: (a) schematic, (b) OAM wavefront, and (c) fabricated device. (d,e) Broadband OAM laser using a tapered structure [53]: (d) schematic and (e) fabricated device. Figure adapted from (a-c), ref. [52], AAAS; and (d,e), ref. [53], with permission.

# IV. CHIRAL PROPAGATION

The breaking of mirror symmetry between CW and CCW propagating modes directly allows chirality for another optical quantity: the chiral evolution of canonical momentum in the system [54-57]. As discussed by Miller [58], the realization of the directionality in photonics is crucial for an input-output isolation [59] in optical logic devices. Although it is well known that PT symmetry cannot break the Lorentz reciprocity [60,61], the asymmetric mode conversions and chiral evolution of wave propagation in complex potentials have enabled various optical functionalities for the routing of wave flows.

From the use of complex potentials composed of gain and loss media, PT-symmetric systems provide a suitable platform for the directional emission of a laser [54]. By asymmetrically imposing the scatterers on a whispering-gallery-mode (WGM) resonator, Peng *et al.* showed that the symmetry between CW and CCW propagation is broken, achieving directional lasing [56] (Fig. 4a). From the time-reversal symmetry relation between lasing and absorbing mechanisms [62-64], the mirror symmetry breaking for the CW and CCW has also been extended to the realization of the chiral absorber for guided-wave structures (Fig. 4b,c) [57].

The directional property obtained from PT-symmetric chirality was also applied to sensing applications suggested by Wiersig [65,66], which was successfully demonstrated by an EP-based nanoparticle sensor that detects the perturbation from the EP [67]. In this EP-based sensor, the frequency splitting through the emergence of a nanoparticle is proportional to the square root of the perturbation, as shown in the term $(\kappa^2 - V_i^2)^{1/2}$ in Eq. (2) (Fig. 4d), which is more sensitive than the linear dependency of the traditional sensing methodology through the diabolic point [68]. This high sensitivity of the EP dynamics is maintained for larger intrinsic backscattering of the unperturbed non-Hermitian system than the backscattering induced by the perturbation (Fig. 4e) [66,67]. This EP sensor is therefore appropriate for detecting sufficiently small perturbations, such as nanoscale sensing problems.

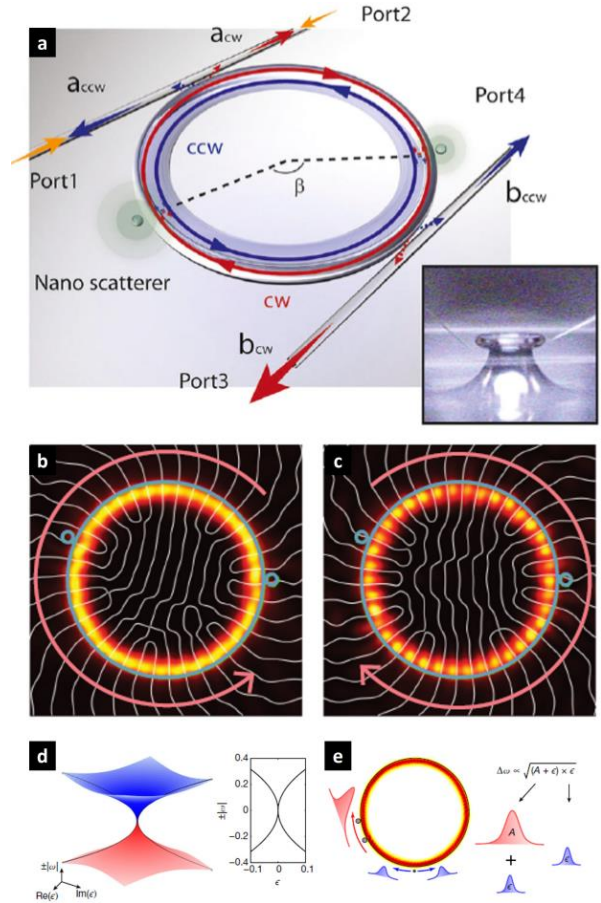

Figure 4. Chiral propagations in non-Hermitian potentials. (a) A schematic of a WGM resonator for observing the chiral wave propagation [56]. (b,c) Asymmetric wave propagation for chiral absorption: (b) CCW and (c) CW wave propagation [57]. (d,e) Operation principle of the EP sensor: (d) square root response and (e) the physical origin from the large backscattering induced by the unperturbed system [67]. Figure adapted from (a), ref. [56], NAS; (b,c), ref. [57], APS; and (d,e), ref. [67], Springer Nature, with permission.

For the chirality of optical wavefront and propagation in Sections III and IV constructed on multimode systems such as the OAM or azimuthal modal numbers, two theoretical aspects need to be considered, degeneracy and disorder, compared to the chirality in the SAM in Section

II with only two degrees of freedom. First, Ge *et al.* demonstrated that the degeneracy of the system beyond 1D results in the absence of a PT-symmetric transition even for the infinitesimal breakdown of the time operator, thus prohibiting a real spectrum (Fig. 5a,b) [69]. The conservation of some further discrete spatial symmetries is required for the emergence of a PT-symmetric transition with the conservation of a real spectrum over a finite interval (Fig. 5c,d). Second, although most research addresses the chiral eigenmode at the EP defined by PT symmetry, which requires a rigorous spatial symmetric configuration, it was shown that the chiral wave behaviors in non-Hermitian photonics can also be obtained in disordered structures, even with the real eigenvalue (Fig. 5e-g) [70]. Starting from the randomly distributed optical resonators, the inverse design in the reciprocal space derives the necessary distribution of complex on-site energy for the target eigenmode, which enables control over the topological charge in wave evolution. This result has been extended to the design of complex optical potentials for necessary functionalities, such as target decoupling [71] and Bohmian photonics [72] for constant-intensity waves [73-75] and phase-conserved energy confinement.

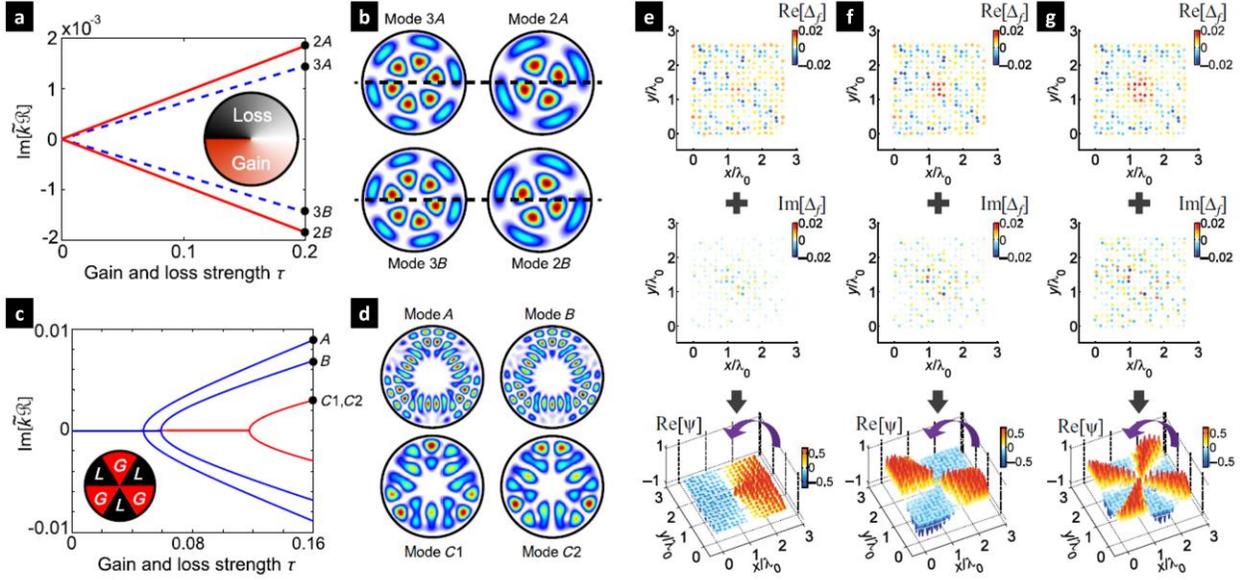

Figure 5. Roles of degeneracy and disorders in chirality in non-Hermitian potentials. (a) The absence of PT-symmetric transitions: (b) modal profiles [69]. (c) The emergence of PT-symmetric transitions with discrete spatial symmetry: (d) modal profiles [69]. (e-g) Realizations of the chiral wave evolution in disordered photonic networks with different topological charges [70]. Figure adapted from (a-d), ref. [69], APS; and (e-g), ref. [70], AAAS, with permission.

## V. CHIRAL GEOMETRIC PHASE

In Sections II-IV, we focused on the chirality of optical quantities defined in real space: the SAM, OAM, and canonical momentum. In this section, we present a more abstract type of chirality in non-Hermitian photonics, namely, chiral phase evolution in an encircling of the EP in the system parameter space. This type of chirality is in line with the well-known concept of "geometric phases" in Hermitian systems, including the Pancharatnam [76], Berry [77], and Aharonov-Anandan phases [78], which denote the observable quantity from gauge-dependent quantities such as vector potentials by introducing the closed path integral around the arbitrary parameter space. The geometric nature of non-Hermitian systems appears in the asymmetric phase evolution of wavefunctions through the encircling of the EP.

The encircling of the EP was first investigated by Heiss *et al.* [32,33,79] in general 2D systems as the local approximation of an isolated EP in infinite-dimensional problems. Because the EP leads to the coalesced eigenvalues through the intersecting complex Riemann sheets, the encircling of the EP forms the topological structures (Fig. 6a,b), which results in the interchange of two eigenmodes, while only one of them acquires a geometric phase as the "chiral geometric phase". This expectation in abstract models has been confirmed with dissipative resonators [80] and molecular photodissociation [81]. The chiral property of the encircling of the EP was observed in microwave resonators with different linewidths of each uncoupled resonator [82], constituting the EP of the passive PT symmetry with small and large loss components [26].

The encircling of the EP has recently been extended to various fields with different platforms and concepts, for example, microwave [83,84] and optical [85,86] waveguides, optomechanics [87], and anti-PT-symmetric systems [88]. These efforts have provided device applications in photonics, especially for directionality. In terms of device functionalities, the realization of EP-encircling through the control of waveguide coupling and the wavevector of a single waveguide enables the strong attenuation of one of two transverse modes, achieving asymmetric mode switching (Fig. 6c) [83]. This asymmetric nature also allows the realization of an optical omnipolarizer [85] and non-reciprocal energy transfer between vibrational modes [87]. Recently, important theoretical discoveries have been achieved in this emerging field, such as the achiral property with a starting point of the encircling at the broken PT symmetry [84], the edge state in a non-Hermitian lattice from the EP-encircling in the momentum space [89], the general formulation of describing arbitrary loops enclosing multiple EPs [90], and the recent extension to the EP from anti-PT symmetry [88]. The practical demonstration of dynamical encircling was also achieved in the silicon photonics platform, which enables the practical realization of non-Hermitian photonics with integrated photonic structures (Fig. 6d) [86].

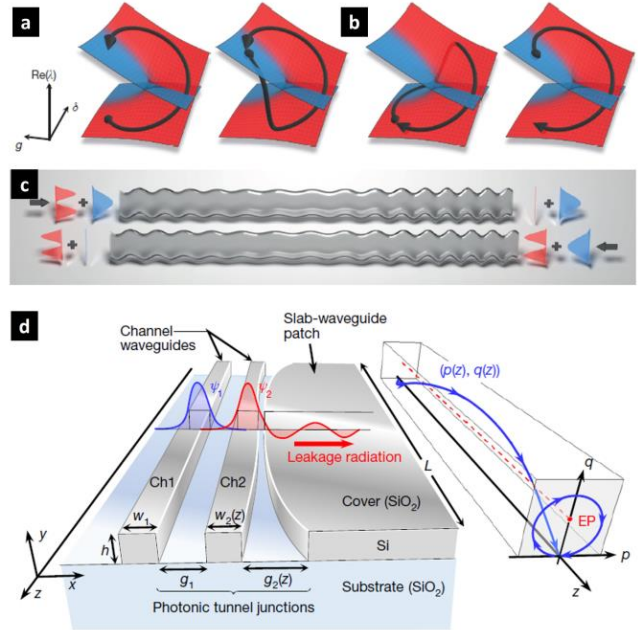

Figure 6. Chiral geometric phase in non-Hermitian potentials. (a,b) Chiral encircling around the EP [83]: (a) Evolution of two eigenmodes with starting points on different Riemann sheets for a CCW loop and (b) the same as that for a CW loop. (c) Asymmetric mode switching based on the dynamical encircling [83]. (d) Silicon photonics platform for the encircling of the EP in the on-chip photonic device [86]. Figure adapted from (a-c), ref. [83], Springer Nature; and (d), ref. [86], Springer Nature, with permission.

## VI. CONCLUSION AND OUTLOOK

The singular form of chirality in non-Hermitian wave systems, which is distinct from conventional optical chirality, has enabled the expansion of the role of chirality not only in photonics but also in general wave physics, such as quantum mechanics, acoustics, and optomechanics. In comparison with different phase velocities or absorptions of circular polarizations in traditional chirality, non-Hermitian systems have stimulated generalized definitions of chirality for waves, achieving exotic wave phenomena: EP-induced chiral convergence of wave quantities (Sections II-IV) and geometric and topological properties distinct from those in Hermitian systems (Section V). According to its abstract form protected by PT symmetry, non-Hermitian chirality can be implemented in arbitrary wave quantities, namely, the SAM, OAM, canonical momentum, and geometric phase. These generalities have been directly applied to practical device applications as discussed in this review, namely, universal polarizers, OAM lasers, directional lasing, chiral absorbers, nanoparticle sensors, and asymmetric modal switching.

By constructing the connections to the recent emerging concepts and phenomena for new degrees of freedom in photonics, the range of the non-Hermitian chirality will be further extended. For example, we envisage the extension of non-Hermitian chirality to novel wave quantities and platforms: (i) non-Hermitian chirality in newly synthetized wave quantities by utilizing synthetic dimensions [91,92] for the physical axes for PT symmetry, (ii) transverse spin [43,93] and its generalization to 3D localized fields for the full realization of polarization states in the meridional plane, (iii) topological properties [11,94,95] with their non-Hermitian extensions for error-robust active photonic devices, (iv) disordered photonics [96-98] for the symmetry-broken realization of non-Hermitian chirality with the independent control of optical amplitude and phase [72], and (v) supersymmetry optics [99-103] for non-PT-symmetric non-Hermitian chirality especially for multilevel OAM devices.

## ACKNOWLEDGMENT

We acknowledge financial support from the National Research Foundation of Korea (NRF) through the Global Frontier Program (2014M3A6B3063708), the Basic Science Research

Program (2016R1A6A3A04009723), and the Korea Research Fellowship Program (2016H1D3A1938069), which are all funded by the Korean government.# REFERENCES

1. M. K. Gaillard, P. D. Grannis, and F. J. Sciulli, "The standard model of particle physics," Rev. Mod. Phys. **71**, S96 (1999).

2. K. Soai, T. Shibata, H. Morioka, and K. Choji, "Asymmetric autocatalysis and amplification of enantiomeric excess of a chiral molecule," Nature **378**, 767 (1995).

3. P. L. Polavarapu, "Optical rotation: recent advances in determining the absolute configuration," Chirality **14**, 768-781 (2002).

4. Y. Tang, and A. E. Cohen, "Optical Chirality and Its Interaction with Matter," Phys. Rev. Lett. **104**, 163901 (2010).

5. K. Y. Bliokh, and F. Nori, "Characterizing optical chirality," Phys. Rev. A **83** (2011).

6. A. Serdyukov, I. Semchenko, S. Tretyakov, and A. Sihvola, *Electromagnetics of bi-anisotropic materials: Theory and applications* (Gordon and Breach Science, 2001).

7. A. Papakostas, A. Potts, D. M. Bagnall, S. L. Prosvirnin, H. J. Coles, and N. I. Zheludev, "Optical Manifestations of Planar Chirality," Phys. Rev. Lett. **90**, 107404 (2003).

8. Y. Tang, and A. E. Cohen, "Enhanced enantioselectivity in excitation of chiral molecules by superchiral light," Science **332**, 333-336 (2011).

9. J. B. Pendry, "A chiral route to negative refraction," Science **306**, 1353-1355 (2004).

10. X. Piao, S. Yu, J. Hong, and N. Park, "Spectral separation of optical spin based on antisymmetric Fano resonances," Sci. Rep. **5**, 16585 (2015).

11. A. B. Khanikaev, S. H. Mousavi, W. K. Tse, M. Kargarian, A. H. MacDonald, and G. Shvets, "Photonic topological insulators," Nat. Mater. **12**, 233-239 (2013).17